\documentclass[12pt]{iopart} 

\usepackage[T1]{fontenc}
\usepackage[english]{babel}
\usepackage[utf8]{inputenc} 
\usepackage{graphicx}
\usepackage{mathcmd}
\usepackage{enumerate}
\usepackage{subfig}

\newcommand{\im}[2][]{\includegraphics[#1]{Figures/#2}}

\begin{document}
\title{Structural properties of hard disks in a narrow tube}
\author{S Varga$^1$, G Balló$^2$ and P Gurin$^1$}
\address{$^1$ Institute of Physics and Mechatronics, University of Pannonia, PO Box 158, H-8201 Veszprém, Hungary}
\address{$^2$ Department of Electrical Engineering and Information Systems, University of Pannonia, Egyetem u. 11, H-8201 Veszprém, Hungary}
\eads{\mailto{vargasz@almos.vein.hu}, \mailto{ballo.g@virt.uni-pannon.hu}, \mailto{gurin@almos.vein.hu}}

\begin{abstract}
Positional ordering of a two-dimensional fluid of hard disks is examined in such narrow tubes where only the nearest-neighbor interactions take place. Using the exact transfer-matrix method the transverse and longitudinal pressure components and the correlation function are determined numerically. Fluid-solid phase transition does not occur even in the widest tube, where the method just loses its exactness, but the appearance of the dramatic change in the equation of state and the longitudinal correlation function shows that the system undergoes a structural change from a fluid to a solid-like order. The pressure components show that the collisions are dominantly longitudinal at low densities, while they are transverse in the vicinity of close packing density. The transverse correlation function shows that the size of solid-like domains grows exponentially with increasing pressure and the correlation length diverges at close packing. It is managed to find an analytically solvable model by expanding the contact distance up to first order. The approximate model, which corresponds to the system of hard parallel rhombuses, behaves very similarly to the system of hard disks.
\end{abstract} 

\noindent{\it Keywords\/}: confined systems, positional order, transfer matrix method, exactly solvable models

\maketitle

\section{Introduction}
\label{sec:intro}
The knowledge of the thermodynamics of fluids in extreme confinements has a great 
importance due to the wide range of chemical and biological applications and the appearance of
nanoscale-fluids in nature \cite{gelb}. The most important examples are the micro- and nanofluidic devices
\cite{romanowsky,smith}, carbon nanotubes \cite{majumde,kyakuno}, biological ion-channels \cite{boda} and adsorption in porous materials such
as the zeolite \cite{rutkai}. To capture the essence of the rather complex behaviour of such confined systems
it is important to apply some simplifying assumptions for the particle-particle and particle-confining
wall interactions. Using hard body molecular models and hard confining walls it is possible to study
purely the geometrical packing effects as the energetic effect are not included in these models. In this
regard much attention has been paid to understand the structure and the phase behaviour of several
confined systems such as the ordering of hard spheres in narrow cylindrical pore \cite{gordillo,duran,mughal} and slit
pores \cite{kim} and the helical packing of soft spheres in nanotubes \cite{pickett,lohr}. The rapid development of
the X-ray techniques makes also possible to study extremely confined fluids, where the particles are
confined between two hard walls with thickness just slightly higher than the diameter of the particle
\cite{nygard}. These systems can be often modeled as quasi-one-dimensional systems.

Since the seminal work of Tonks \cite{tonks} considerable progress has been made to understand the
thermodynamic and structural properties of one-dimensional systems. The pair correlation function
was determined in various systems interacting with short-range pair potentials \cite{salsburg,vo,fantoni}, the link
between the short-ranged pair potential and the pair correlation functions was examined in a charged-
stabilized colloidal system trapped to one dimension \cite{hansen} and the long-range interaction induced
freezing transition of hard rods was studied by \cite{carraro}. Even our knowledge of the ordering properties
of one-dimensional hard rod fluid (Tonks gas) has been widened by the exact determination of the
percolation size \cite{drory} and by the study of the residual multiparticle entropy at the onset of solidlike
behaviour \cite{giaquinta}.

The examination of confined two-dimensional (2D) hard particle fluids can be considered as
a natural extension of the one-dimensional studies, if the available space for the particles is infinitely
long in one dimension, but it is very restricted in other dimension. This condition guarantees that
the phase behaviour of the confined fluid does not deviate substantially from that of Tonks gas. Of
course, with weakening confinement the system gets naturally closer to the two-dimensional bulk
fluid and the thermodynamic properties of such fluids get very far from that of Tonks gas \cite{jaster,mak}.
Several studies are devoted to extend the exact Tonks theory of one-dimensional fluid to the confined
two-dimensional one \cite{barker,post,kofke,schwartz}. If the confinement and the pair interaction does allow only nearest
neighbor interactions, it is possible to derive such an integral equation which provides the exact
thermodynamics of the system. In the system of 2D hard disks confined into a narrow tube, the
equation of state has been determined by the iterative solution of the corresponding eigenvalue
equation \cite{kofke}. Later analytical equation has been derived for the equation of state in the high pressure
limit, which is not valid at low densities \cite{percus,kamenetskiy}. Recently both the transverse and the longitudinal pressure components of the hard and soft disk have been determined at such low densities, where the second
virial theory is still valid \cite{forster,mukamel}. In this paper we extend the previous studies to examine the
local structure of the hard disk fluid at both low and high pressures and we devise an analytically
solvable model system which follows closely the phase behaviour of the hard disks. We calculate the
transverse density profile, the pair correlation function and the longitudinal pair distribution function
to examine the growth of the solid-like clusters with increasing pressure. We pay special attention to
the effects of the structural change from a Tonks gas behaviour to a solid-like one.

Our paper is organized as follows: the transfer matrix method is presented for the confined
hard disks in the next section. The eigenvalue equation is derived for the determination of the
Gibbs free energy, the equation of state, the correlation function and the local density. Moreover
we present the details of our MC simulation study for the pair correlation function measured in the
direction of the tube’ s long axis. In section \ref{sec:analytic} we show that analytical results can be derived for both
thermodynamic and structural properties if the contact distance is expanded up to the first order.
The results are shown to be exact for the system of parallel hard rhombuses placed into a very narrow
tube, but it is approximate for hard disks even in the vicinity of the close packing limit. We compare
our numerical, analytical and MC results and we also examine the fluid-solid like structural change,
which takes place with increasing density in section \ref{sec:results}. Conclusions are given in section \ref{sec:concl}.

\section{Transfer matrix method of the confined two-dimensional hard disks}
\label{sec:tfmat}
\begin{figure*}
\begin{center}
\im[width=0.6\textwidth]{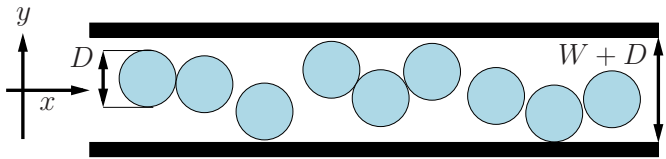}
\end{center}
\caption{Schematic representation of the system of hard disks confined between two hard walls, where $D$ is the diameter of the disks and $W+D$ is the wall-to-wall separation. Overlaps are not allowed between the disks and the vertical position of the disks is confined in the interval of $-W/2<y<W/2$.\label{fig:1}}
\end{figure*}

We consider an isobaric ensemble of $N$ freely moving two-dimensional hard disks in such
a restricted geometry where only nearest neighbor interactions are allowed (see figure \ref{fig:1}).
The interaction potential between two neighboring hard disks is defined by
\begin{equation}
\label{eq:1}
u(r_{i,i+1})=
\left\lbrace
\begin{array}{c}
\infty,\ r_{i,i+1}\leq D\\
0,\ r_{i,i+1}>D
\end{array} 
\right.\ ,
\end{equation}
where $r_{i,i+1}=|\bi{r}_{i+1}-\bi{r}_i|$ is the length of the centre-to-centre distance vector between particles $i$ and $i+1$, $\bi{r}_i=(x_i,y_i)$, $x_i$ and $y_i$ denote cartesian coordinates of the position vector of the particle $i$ and $D$ is the diameter of the disk. Equation \eref{eq:1} precludes the overlap between the particles. To restrict the number
of interactions only to the nearest neighbors a confining external field is applied, which acts only on
the $y$ coordinates of the particle's centers and allows the particles to stay in a very narrow tube such that
\begin{equation}
\label{eq:2}
u_{\mathrm{ext}}(y_i)=
\left\lbrace
\begin{array}{c}
0,\ |y_i|\leq W/2\\
\infty,\ |y_i|>W/2
\end{array} 
\right.\ ,
\end{equation}
where $W$ is practically the width of the tube even if there is an additional $D$ distance in the tube
which is inaccessible for the centers of the particles (see figure 1). It is easy to show that the tube
width must be in the interval $0<W<D\sqrt{3}/2$ to exclude the next-nearest neighbor interactions. In
such a narrow tube the system is anisotropic, i.e. the horizontal and vertical pressure components $P_{xx}$ and $P_{yy}$
are different. From the two possible options to define the Gibbs free energy ($G$) from the 
Helmholtz free energy ($F$) we choose the horizontal pressure component to get $G$ from the Legendre transformation, i.e. $G=F+P_{xx}A$, where $A=LW$
is the accessible area for the centre of the particle and $L$ is the length of the system along the $x$ axis. The corresponding isobaric partition function of
the confined system can be written as
\begin{eqnarray}
\label{eq:3}
\fl
Z_{NP_{xx}T}=\frac{1}{\Lambda^{2N}}\int\limits_0^\infty\dd x_1\cdots\dd x_N\int\limits_{-W/2}^{W/2}\dd y_1\cdots\dd y_N 
\exp\left(-\beta\sum_{i=1}^N\left(u(r_{i,i+1})+P_{xx}Wx_{i,i+1}\right)\right) ,
\end{eqnarray}
where $\beta=1/k_\mn{B}T$ ($T$ being the temperature and $k_\mn{B}$ the Boltzmann constant) and $\Lambda=h/\sqrt{2\pi k_\mn{B}Tm}$ is the thermal de Broglie wavelength. Note that the missing $N!$ term in the partition function is due to the use of the fixed $x_1\leq x_2 \leq \cdots \leq x_N$ order. This order can be fixed because the tube is enough narrow. Furthermore the finite integral in $x_i$ is replaced by the integral with infinite upper bound in the isobaric ensemble and we use the fact that $L=\sum_{i=1}^Nx_{i,i+1}$ where $x_{i,i+1}=x_{i+1}-x_i$. Changing from the integration variable $x_i$ to $x_{i,i+1}$ and using equation \eref{eq:1} it is easy to prove that
\begin{eqnarray}
\label{eq:4}
\fl
Z_{NP_{xx}T}=\frac{1}{\Lambda^{2N}}\int\limits_{-W/2}^{W/2}\dd y_1\cdots\dd y_N \nonumber \\ 
\int\limits_{\sigma(y_1,y_2)}^{\infty}\dd x_{1,2}\;\exp\left(-\beta P_{xx}Wx_{1,2}\right)\;\cdots
\int\limits_{\sigma(y_N,y_1)}^{\infty}\dd x_{N,1}\;\exp\left(-\beta P_{xx}Wx_{N,1}\right) , 
\end{eqnarray}
where $\sigma(y_i,y_{i+1})=\sqrt{D^2-(y_i-y_{i+1})^2}$ is the horizontal projection of the contact distance between two hard disks being at the positions $y_i$ and $y_{i+1}$. The lower bound of the integrals in $x_i$ comes from the overlap condition, which prescribes that $r_{i,i+1}\geq D$. We have exploited the periodic boundary condition in the $x$ direction, too, which means that $r_{N+1}=r_1$. Now we are in the position to perform the positional integrations in $x_{i,i+1}$ analytically, i.e. 
\begin{equation}
\label{eq:5}
Z_{NP_{xx}T}=\left(\frac{D}{\Lambda^{2}}\right)^N\int\limits_{-W/2}^{W/2}\dd y_1\cdots\dd y_N K(y_1,y_2)K(y_2,y_3)\cdots K(y_N,y_1)\ ,
\end{equation}
where $K(y_i,y_{i+1})=\exp(-\beta P_{xx}W\sigma(y_i,y_{i+1}))/\beta P_{xx}WD$ can be considered as an element of an infinite-dimensional matrix ($\hat{K}$). In this formalism the matrix product is given by $K^2(y_i,y_{i+2})=\int\limits_{-W/2}^{W/2}\dd y_{i+1}K(y_i,y_{i+1})K(y_{i+1},y_{i+2})$. Therefore the partition function can be considered as the trace of the matrix $\hat{K}^N$, that is $Z_{NP_{xx}T}=\trace{\hat{K}^N}=\int\limits_{-W/2}^{W/2}\dd y K^N(y,y)$. Due to the fact that the result of trace operation is independent of the used basis, it is most convenient to evaluate the partition function in the eigenfunction basis where the matrix $\hat{K}$ is diagonal. Denoting the eigenvalues of $\hat{K}$ in decreasing order in the absolute value as $|\lambda_0|>|\lambda_1|>|\lambda_2|\dots$, equation \eref{eq:5} reduces to $Z_{NP_{xx}T}=\sum_{k=0}^\infty\lambda_k^N$, where the eigenvalues are the solutions of the following eigenvalue equations
\begin{equation}
\label{eq:6}
\int\limits_{-W/2}^{W/2}\dd y'K(y,y')\psi_k(y')=\lambda_k\psi_k(y)\ ,
\end{equation}
where $\psi_k(y)$ is the $k$th eigenfunction ($k=0,1,2,\dots$) and $K(y,y')$ is the integral kernel defined in equation \eref{eq:5}. To our best knowledge, the above eigenvalue equation does not have analytical solutions for hard disks. Therefore we solve it by the iteration using the trapezoidal quadrature for the numerical integration. We can obtain the working form of equation \eref{eq:6} by the use of the normalization constraint of the eigenfunctions, that is $\int\limits_{-W/2}^{W/2}\dd y\psi_k(y)^2=1$. In this way we write the equations of the eigenfunctions and the corresponding eigenvalues as follows
\begin{equation}
\label{eq:7}
\psi_k(y)=\frac{\int\limits_{-W/2}^{W/2}\dd y'K(y,y')\psi_k(y')}{\int\limits_{-W/2}^{W/2}\dd y'\dd y''K(y',y'')\psi_k(y'')\psi_k(y')}\ ,
\end{equation}
and
\begin{equation}
\label{eq:8}
\lambda_k=\int\limits_{-W/2}^{W/2}\dd y\dd y'K(y,y')\psi_k(y')\psi_k(y)\ .
\end{equation}
In the thermodynamic limit ($N\rightarrow\infty$) the largest eigenvalue ($\lambda_0$) has the most significant contribution to the Gibbs partition function, i.e. the Gibbs free energy takes the following form: $\beta G/N=-\ln(\lambda_0/D)$, where the constant term proportional to $\ln(\Lambda/D)$ is omitted. The equation of state can be obtained form the standard thermodynamic relation between the Gibbs free energy and the pressure, which is now $W/\rho=\partial g/\partial P_{xx}$, where $\rho=N/L$ is the linear density and $g=G/N$. Similarly the transverse pressure can be also derived from the Gibbs free energy as follows
\begin{equation}
\label{eq:9}
P_{yy}=P_{xx}-\rho\pdifn{g}{W}\Bigg|_{P_{xx}}\ .
\end{equation}

Regarding the positional distributions of the disks, $\psi_0(y)$ eigenfunction gives account of the positional ordering along the vertical direction, because $\psi_0^2(y)$ is the probability density distribution function and $\rho(y)=\rho\psi_0^2(y)$ is the particle number density distribution function along $y$ axis. Other important function which measures the posititional correlation in the vertical directions between two particles $i$ and $j$ is defined as $g(i,j)=\Big\langle\big(y_i-\langle y_i\rangle\big)\big(y_j-\langle y_j\rangle\big)\Big\rangle$. For the present case we only need to calculate the $NP_{xx}T$ ensemble average of the products of the horizontal positions of particles $i$ and $j$ ($g(i,j)=\langle y_iy_j\rangle$), because the average position of the particles is trivially zero ($\langle y_i\rangle=0$). It can be shown by using the eigenvalues and eigenfunctions that
\begin{equation}
\label{eq:10}
g(i,j)=\sum_{k=1}^\infty\left(\frac{\lambda_k}{\lambda_0}\right)^{|i-j|}\left[\int\limits_{-W/2}^{W/2}\dd y\;\psi_0(y)\,y\,\psi_k(y)\right]^2\ .
\end{equation}
Note that the fluctuation in the vertical position is given by
\begin{equation}
\label{eq:11}
g(i,i)=\int\limits_{-W/2}^{W/2}\dd y\;\psi_0^2(y)\,y^2\ .
\end{equation}
It can be seen from equation \eref{eq:10} that the correlation function decays exponentially for $|i-j|\rightarrow\infty$, i.e. we can write that $g(i,j)\sim\exp(-|i-j|/\xi)$, where the correlation length ($\xi$) is given by $\xi=(\ln|\lambda_0/\lambda_1|)^{-1}$. Besides the transverse correlations it is also necessary to get information about the longitudinal positional order to characterize the structure of the confined disks. We do this by the MC simulation of the longitudinal pair distribution function.  This function is computed by our own Monte Carlo simulation in canonical ensemble (NAT) from the expression 
\begin{equation}
\label{eq:12}
g\bigg(x=|x'-x''|\bigg)=\rho^{-2}\left\langle\sum_{i}\sum_{j\neq i}\delta(x_i-x')\,\delta(x_j-x'')\right\rangle_{\mn{NAT}}\ .
\end{equation}
In this expression the configurations are collected from the simulation of $1000$ hard disks, which are sampled from $10^5$ configurations beyond an equilibration process of $10^5$ configurations. The starting configuration is a triangular lattice, where the particles are at the two walls in alternating up-down sequence. In addition, the periodic boundary condition is used in the horizontal direction in all studied cases. Using the Metropolis sampling, the particles are chosen and moved randomly with an adjusted maximum displacement to ensure the requirement of $40$-$50\%$ acceptance probability of all trials. Due to the simple form of the pair potential and the external field, one trial move of a chosen particle is accepted if the particle does not overlap with its neighbors and the walls. 

\section{Analytically solvable model}
\label{sec:analytic}
In this section we proceed by searching for the analytical solution of the eigenvalue equation \eref{eq:6}. We are facing an integral equation which corresponds to a homogeneous Fredholm equation of the second type. Due the non-separable form of the contact distance $\sigma(y_i,y_{i+1})$ and the integral kernel, it is well-known that the Fredholm equation does not have a general solution. In previous studies, Kofke and Post \cite{kofke} solved equation \eref{eq:6} by the standard iterative solution, while Kamenetskiy et al. \cite{kamenetskiy} considered only the high pressure limit and performed a perturbative calculation.  
   
\begin{figure*}
\begin{center}
\im[width=0.6\textwidth]{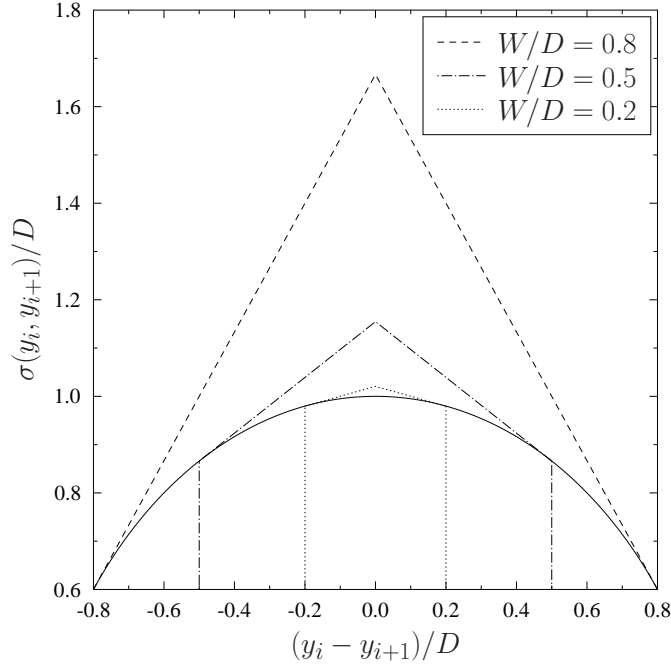}
\end{center}
\caption{Horizontal projection of the contact distance between two disks as a function of their vertical distance. The first order approximation of the excluded distance is shown for different tube widths.\label{fig:2}}
\end{figure*}

Here we show that equation \eref{eq:6} has analytical solutions in a closed form if the horizontal projection of the contact distance is given in the following form 
\begin{equation}
\label{eq:13}
\sigma(y_i,y_{i+1})=c_0+c|y_i-y_{i+1}|\ ,
\end{equation}
where $c_0$ and $c$ are constants. This formula can be seen as an approximative expression for the contact distance of confined hard disks. Expanding the exact contact distance of disks around $y_i-y_{i+1}=\pm W$ up to first order in $y_i-y_{i+1}$ we obtain 
\begin{equation}
\label{eq:14}
c_0=\frac{D^2}{\sqrt{D^2-W^2}}\ ,\qquad c=-\frac{W}{\sqrt{D^2-W^2}}\ .
\end{equation}
The reliability of the first order Taylor expansion is presented in figure \ref{fig:2}, where the curve of the exact contact distance is compared to the approximate formula \eref{eq:13} at different widths of the tube. 
We can see that the equation \eref{eq:13} works quite accurately for narrow tubes, but the error of \eref{eq:13} can even exceed $60\%$ for $W=0.8D$ at zero distance. However, at close packing limit this expansion is justified even at wider tubes because the neighboring disks are forced to be at the opposite walls to fill in the space most efficiently. Furthermore, equation \eref{eq:13} gives exact formula for the hard rhombuses which can not rotate and one of their diagonal axes is constrained to be parallel to the wall. Denoting the lengths of the transverse and longitudinal axes of a rhombus as $a$ and $b$ respectively, it is easy to show that $\sigma=b-b|y_i-y_{i+1}|/a$. Now we are at the stage to present the solutions of equation \eref{eq:6}, which is now supplemented by equation \eref{eq:13} with $c_0$ and $c<0$ constants. The largest eigenvalue and the corresponding eigenfunction are given by
\numparts
\begin{eqnarray}
\label{eq:15}
\lambda_0=D\frac{2c}{(cP^{*})^2-\kappa_0^2}\exp\left(-\frac{c_0P^*}{D}\right),\\ 
\psi_0(y)=\frac{1}{\sqrt{D}}\left(\frac{2\kappa_0}{\sinh(\kappa_0W/D)+\kappa_0W/D}\right)^{1/2}\cosh\left(\kappa_0\frac{y}{D}\right),
\end{eqnarray}
where $\kappa_0$ is the unique positive solution of the following equation
\begin{equation}
\kappa_0=-cP^*\coth\left(\kappa_0\frac{W}{2D}\right) \nonumber\ ,
\end{equation}
\endnumparts
and $P^*=\beta P_{xx}WD$ is the dimensionless preassure.
Using the above eigenvalue we can determine the Gibbs free energy, linear density and the transverse pressure component, while the transverse density profile can be obtained with the aid of $\psi_0$ eigenfunction. It is important to know the other solutions of the eigenvalue equation, because the pair correlation function \eref{eq:10} contains them. In absolute value the second largest eigenvalue, $\lambda_1$, and the corresponding eigenfunction, $\psi_1$, depend on the pressure. In the high pressure case, when $-cP^*W/D>2$,
\numparts
\begin{eqnarray}
\label{eq:16}
\lambda_1=D\frac{2c}{(cP^{*})^2-\kappa_1^2}\exp\left(-\frac{c_0P^*}{D}\right),\\ 
\psi_1(y)=\frac{1}{\sqrt{D}}\left(\frac{2\kappa_1}{\sinh(\kappa_1W/D)-\kappa_1W/D}\right)^{1/2}\sinh\left(\kappa_1\frac{y}{D}\right),
\end{eqnarray}
where $\kappa_1$ is the unique positive solution of the following equation
\begin{equation}
\kappa_1=-cP^*\tanh\left(\kappa_1\frac{W}{2D}\right) \nonumber\ .
\end{equation}
\endnumparts
In the low pressure case, when $-cP^*W/D<2$,
\numparts
\begin{eqnarray}
\label{eq:17}
\lambda_1=D\frac{2c}{(cP^{*})^2+\kappa_1^2}\exp\left(-\frac{c_0P^*}{D}\right),\\ 
\psi_1(y)=\frac{1}{\sqrt{D}}\left(\frac{2\kappa_1}{\kappa_1W/D-\sin(\kappa_1W/D)}\right)^{1/2}\sin\left(\kappa_1\frac{y}{D}\right),
\end{eqnarray}
where $\kappa_1$ is the unique solution of the following equation in the interval $\kappa_1\in[0,\pi D/W]$
\begin{equation}
\kappa_1=-cP^*\tan\left(\kappa_1\frac{W}{2D}\right) \nonumber\ .
\end{equation}
\endnumparts
In the case of $-cP^*W/D=2$ the above two groups of formulas (16) and  (17) give the same limit: $\lambda_1=D\,2c\exp\left(-c_0P^*/D\right)/(cP^{*})^2$, and $\psi_1(y)=2\sqrt{3}W^{-3/2}\,y$. We must mention that $\lambda_1<0$ in all cases. This is the reason why the correlation function shows alternating behaviour.

All of the remaining eigenvalues and eigenfunctions can be summarized for odd and even indices as follows.
\numparts
\begin{eqnarray}
\label{eq:18}
\lambda_i=D\frac{2c}{(cP^{*})^2+\kappa_i^2}\exp\left(-\frac{c_0P^*}{D}\right), \qquad i=2,3,\dots \\
\fl
\psi_i(y)=\cases{
\frac{1}{\sqrt{D}}\left(\frac{2\kappa_i}{\kappa_iW/D+\sin(\kappa_iW/D)}\right)^{1/2}\cos\left(\kappa_i\frac{y}{D}\right),& if $i$ is even
\\
\frac{1}{\sqrt{D}}\left(\frac{2\kappa_i}{\kappa_iW/D-\sin(\kappa_iW/D)}\right)^{1/2}\sin\left(\kappa_i\frac{y}{D}\right),& if $i$ is odd
}
\end{eqnarray}
where $\kappa_i$ ($i=2,3,\dots$) are the unique solutions of the following equation in the intervals $\kappa_i\in[(i-1)\pi D/W,i\pi D/W]$
\begin{equation}
\kappa_i=\cases{
+cP^*\cot\left(\kappa_i\frac{W}{2D}\right)\ ,& if $i$ is even \\
-cP^*\tan\left(\kappa_i\frac{W}{2D}\right)\ ,& if $i$ is odd.
}
 \nonumber
\end{equation}
\endnumparts

Finally we mention that equation \eref{eq:6} has also analitic solution for $c>0$, but they are different and have relevance for the system of freely rotating hard needles in one dimension \cite{gurin}.

\section{Results and discussions}
\label{sec:results}
\begin{figure*}
\begin{center}
\subfloat[]
{   
    \im[width=0.6\textwidth]{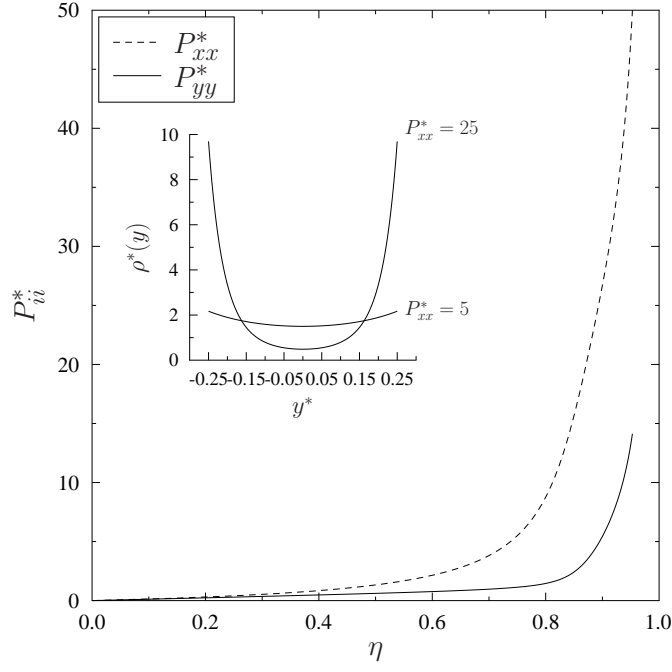}
    \label{fig:3a}
}\\
\subfloat[]
{   
    \im[width=0.6\textwidth]{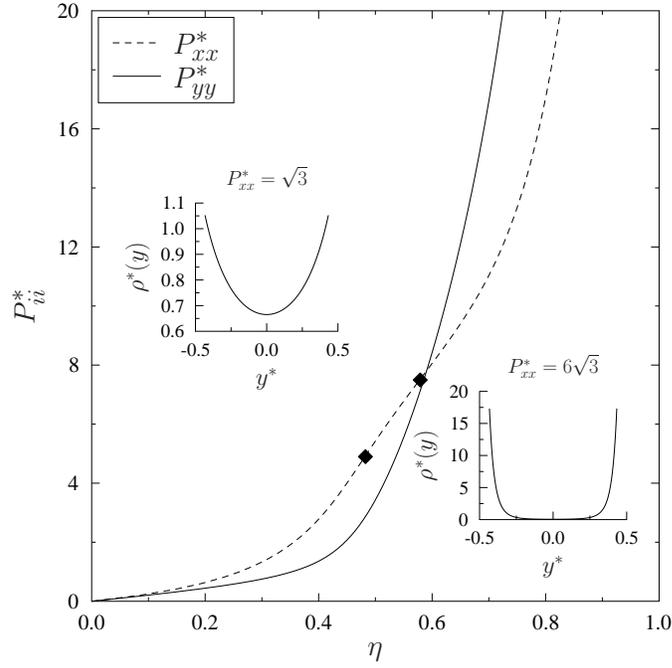}
    \label{fig:3b}
}
\end{center}
\caption{Horizontal ($P_{xx}$) and vertical pressure ($P_{yy}$) components as a function of packing fraction for $W=D/2$ (upper panel) and $W=D\sqrt{3}/2$ (lower panel). The density profiles ($\rho^*(y)=\rho(y)D$) along $y$ axis are shown in the insets. The inflection points are marked by diamond symbols. The pressure components are made dimensionless as follows $P^*_{ii}=\beta P_{ii}DW$, while the packing fraction is defined as $\eta=\rho\sqrt{D^2-W^2}$.\label{fig:3}}
\end{figure*}

\begin{figure*}
\begin{center}
\subfloat[]
{   
    \im[width=0.6\textwidth]{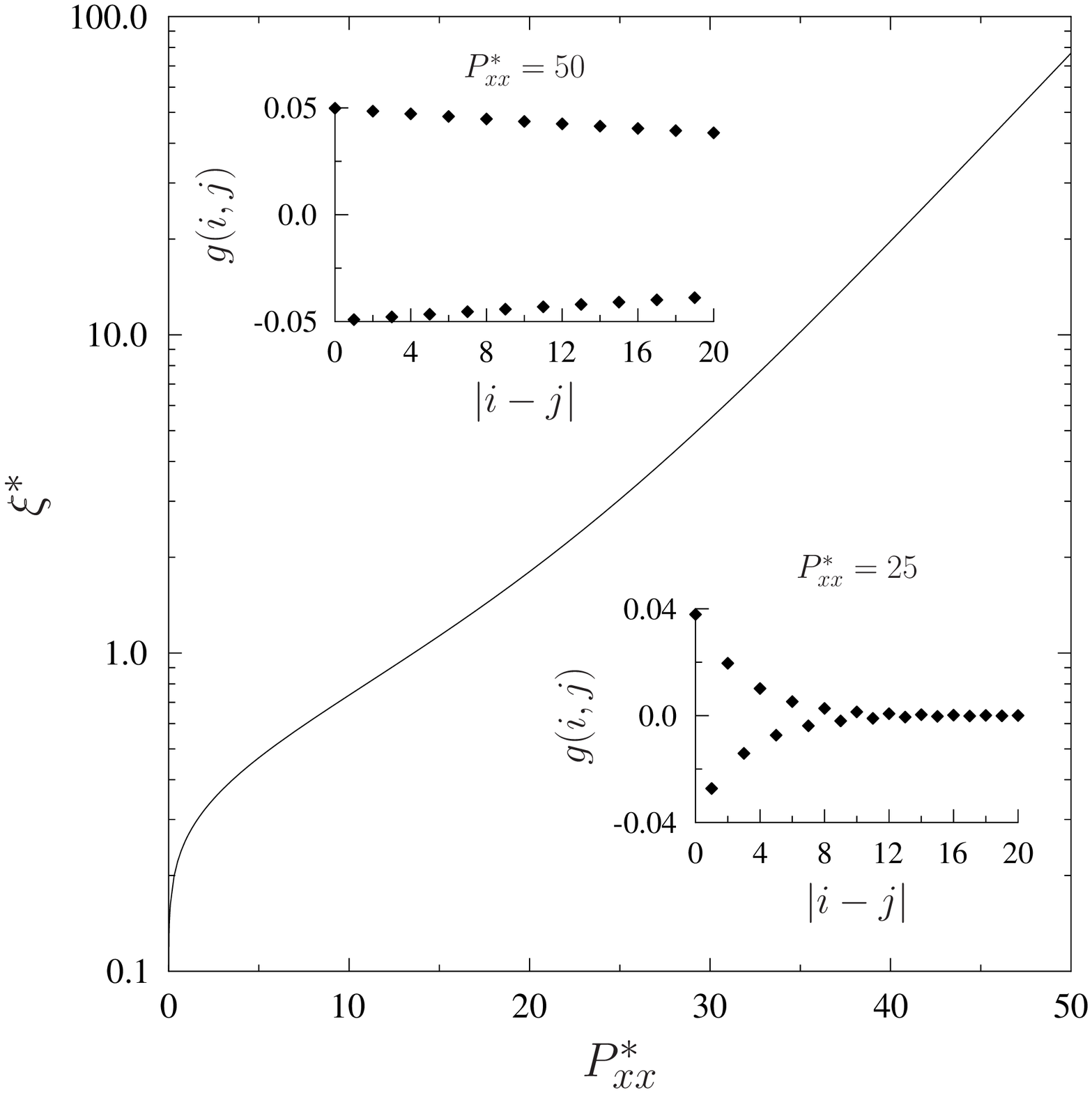}
    \label{fig:4a}
}\\
\subfloat[]
{   
    \im[width=0.6\textwidth]{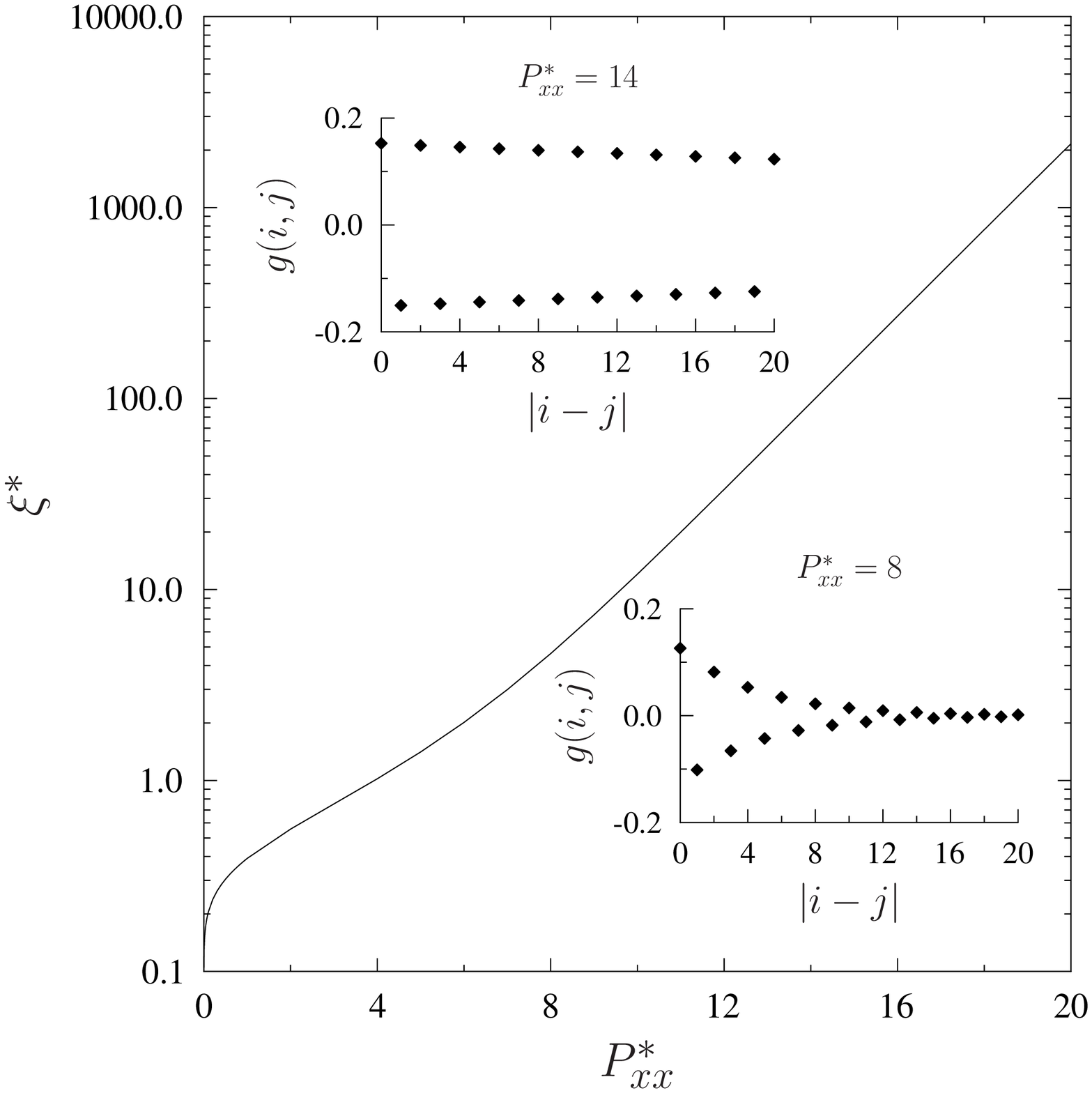}
    \label{fig:4b}
}
\end{center}
\caption{Vertical correlation length ($\xi^*=\xi/D$) as a function of linear pressure ($P^*=\beta P_{xx}WD$) for $W=D/2$ (upper panel) and $W=D\sqrt{3}/2$ (lower panel). Vertical pair correlation function ($g(i,j)=\langle y_iy_j\rangle$) are shown as a function of pair separation ($|i-j|$)  in the insets.\label{fig:4}}
\end{figure*}

In this section we begin with the presentation of our numerical results for the pressure components, density profiles and some structural properties. To do this we make a connection between the present model and an analytically solvable reference system. In the limit of $W\rightarrow 0$, the system becomes a one-dimensional fluid of hard rods with diameter $D$, which is the well-know Tonks gas \cite{tonks}. We can achieve the Tonks gas behaviour at finite width, too, by equating the contact distance to $D$ at any transverse separation, i.e. $\sigma(y_i,y_{i+1})=D$ for all horizontal positions. This case corresponds to the system of parallel hard squares in a tube. It is easy to show that equation \eref{eq:6} gives $\psi_0=1/\sqrt{W}$ and $\lambda_0=\exp(-\beta P_{xx}WD)/(\beta P_{xx}D)$ for the squares and the relation between the pressure and the Gibbs free energy gives back the Tonks equation for the pressure, which is $\beta P_{xx}W=\rho/(1-\rho D)$. From the equation of transverse pressure \eref{eq:9} we get that the squares behaves as an ideal gas between the two walls, i.e. $\beta P_{yy}=N/(LW)$. Since the contact distance deviates only slightly from $D$ in very narrow tubes (see figure \ref{fig:2}), the confined parallel hard squares can considered as a reference system of the hard disks. Our numerical results show that the system of hard disks follows the phase behaviour of hard squares in narrow tubes. This can be observed even in figure \ref{fig:3} for $W=D/2$, where the transverse and longitudinal pressure components are shown as a function of linear packing fraction ($\eta=\rho\sqrt{D^2-W^2}$). 
The longitudinal pressure follows the Tonks equation, while the transverse one can be described by the ideal gas law up to $80\%$ occupation of the longitudinal direction ($\eta\approx0.8$). The transverse density profile also shows that the density is almost constant across the tube for $\beta P_{xx}WD=5$, i.e. $\psi_0=1/\sqrt{W}$ is a reasonable approximation up to $\eta\approx 0.8$ for hard disks. For higher pressures (packing fractions) the density profile is peaked at the walls and there are less particles in the centre of the tube, which is shown at $\beta P_{xx}WD=25$. The results are more interesting at the possible widest tube width ($W=D\sqrt{3}/2$), which can be studied exactly taking into account the nearest neighbour interactions only. The Tonks gas and the ideal gas behaviour are lost at much lower packing and the derivative of the longitudinal pressure shows non-monotonic behaviour with the density. It has two inflection points at intermediate packing fractions and it is crossed by the transverse pressure in the vicinity of larger inflection point. These results suggest that some positional rearrangement must take place in the system to give rise a big change in the pressure. Especially the cross point between the two pressure components is very informative. It means that the particles must interact with each other very strongly along the transverse direction, which can only happen if the particles are in alternating arrangement at the walls. This speculation is also supported by the density profiles shown in the inset of figure \ref{fig:3}. A deeper understanding of the positional ordering can be gained from the transverse pair correlation function \eref{eq:10}, which measures the propagation of the fluctuations between the particles. Figure \ref{fig:4} highlights two important features of the positional ordering. 
Firstly the correlation length diverges with the pressure (density), which shows that the size of the ordered clusters increases and a solid-like structure may emerge at the close packing. Secondly the oscillatory behaviour of the correlation function gives information about the spatial arrangement of the disks, namely if one particle moves into one given transverse direction its neighbor does it in the opposite way. This type of correlation can occur only if the particles are located in an alternating up-down sequence. At low pressures only short range alternating arrangement takes place, but the ordered clusters grows exponentially at higher pressures and the arrangement become similar to a section of a triangular lattice. As the correlation length does not diverge below the close packing, no phase transition between fluid and solid phases can occur in the physical meaningful density range. We can also see that the transverse correlation becomes stronger in wider tubes, as the alternating packing becomes entropically more favored and it is getting harder to make jumps between the two walls.

\begin{figure*}
\begin{center}
\subfloat[]
{   
    \im[width=\textwidth]{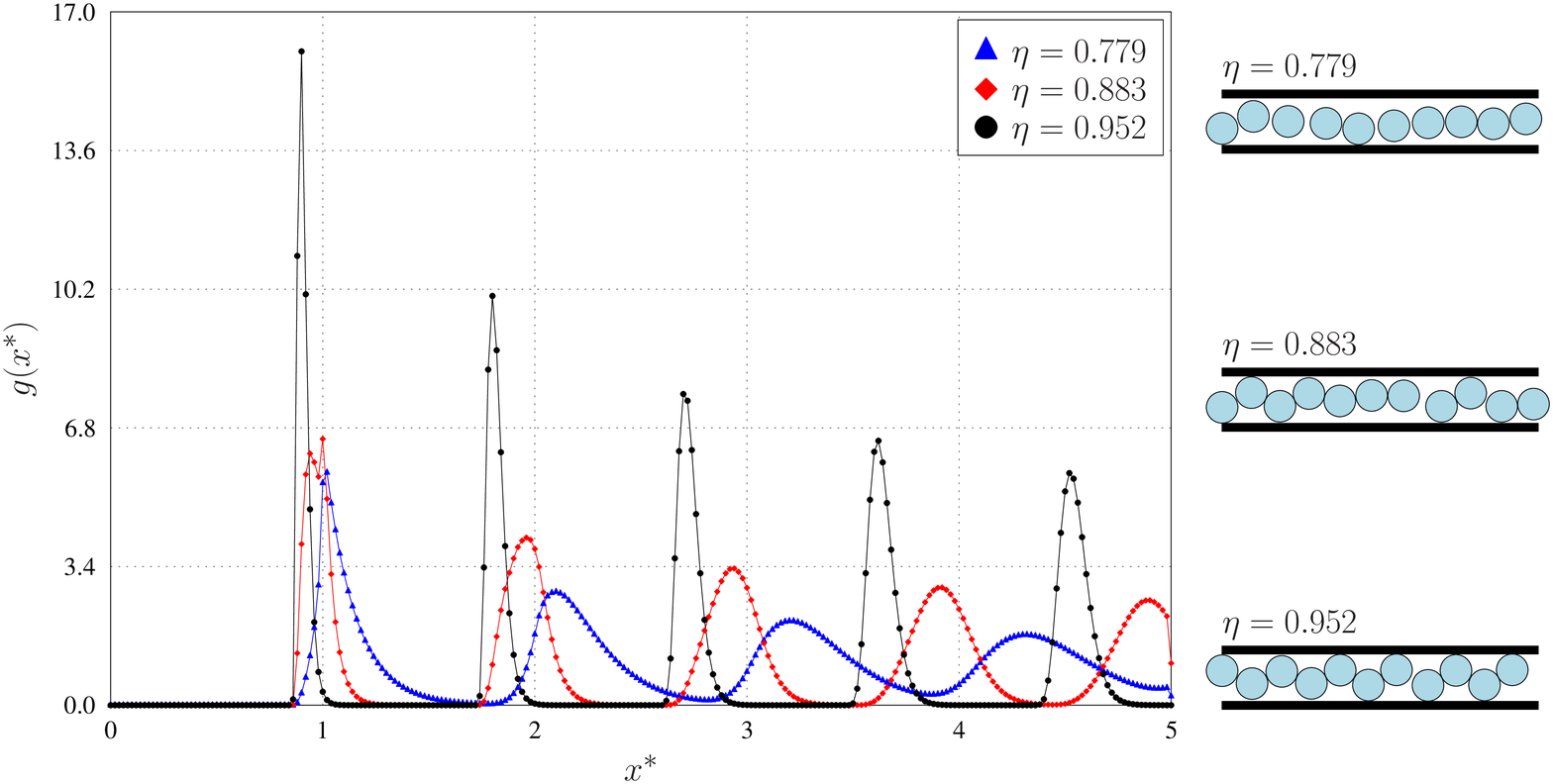}
    \label{fig:5a}
}\\
\subfloat[]
{   
    \im[width=\textwidth]{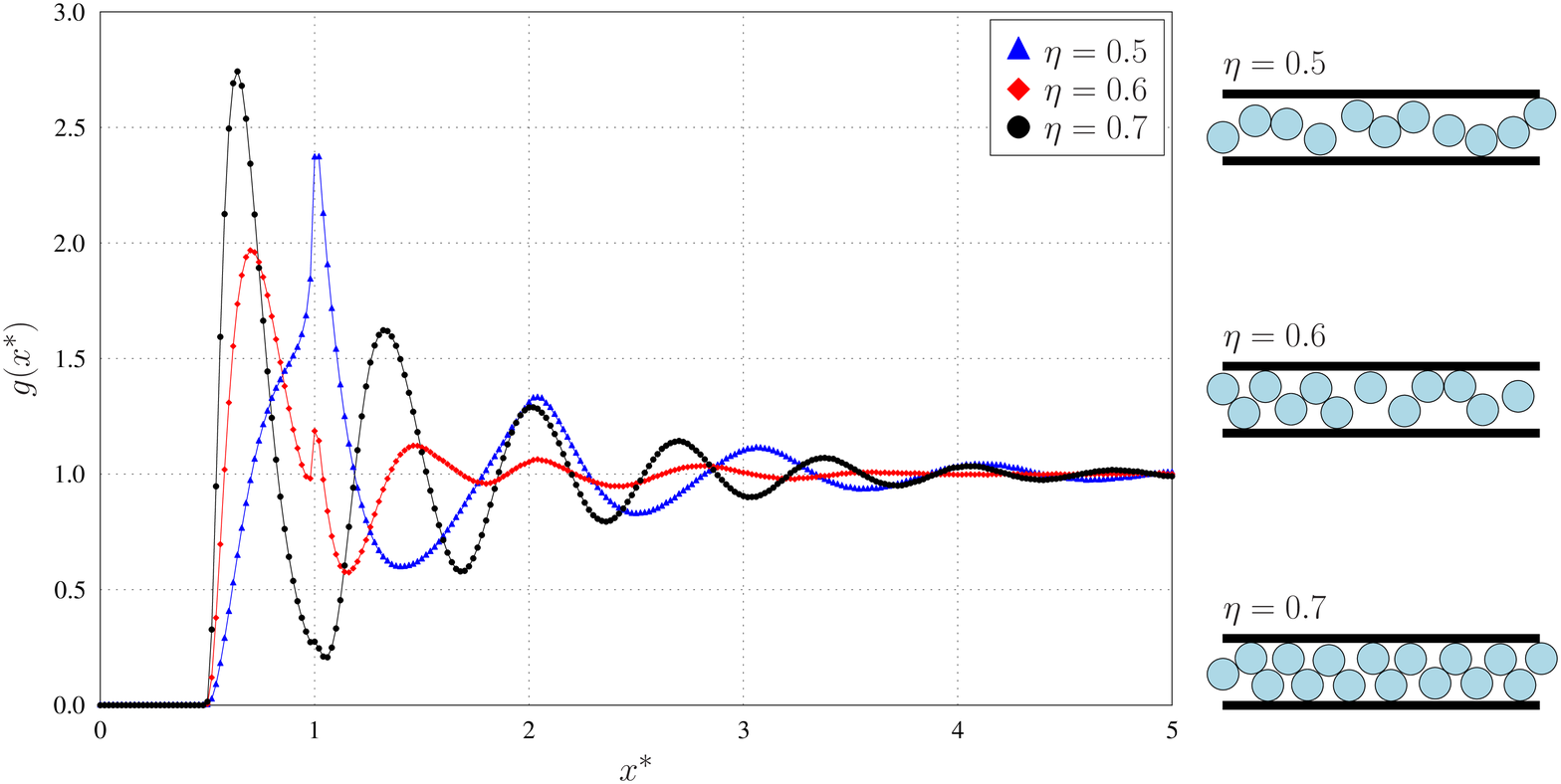}
    \label{fig:5b}
}
\end{center}
\caption{Pair distribution function ($g(x)$) as a function of distance ($x^*=x/D$) for $W=D/2$ (upper panel) and $W=D\sqrt{3}/2$ (lower panel). MC simulation snapshots are shown at the corresponding packing fractions.\label{fig:5}}
\end{figure*}

The longitudinal pair distribution function also helps us to understand the structural changes occurring in the tube. The results of our canonical MC simulation shows that in case of low pressures the average distance between the neighboring disks is in the order of $D$ for both narrow and wide tubes. This means that at low packing the confined hard disk systems behave like the Tonks gas. Figure \ref{fig:5} shows that one additional first peak emerges in the pair distribution function at intermediate packing, which means that the system is just between the Tonks gas and solid-like behaviours. 
In the very dense systems, the second peak completely disappears, the distribution function is extremely peaked and the distance between the peaks is less than $D$. This can only happen if the particles are in the alternating solid-like structure. These results are confirmed by the corresponding simulation snapshots presented in figure \ref{fig:5}, too.  Note that the locations of the inflection points in the transverse pressure are in accordance with the structural changes of the pair distribution function and the simulations snapshots. 

\begin{figure*}
\begin{center}
\im[width=0.6\textwidth]{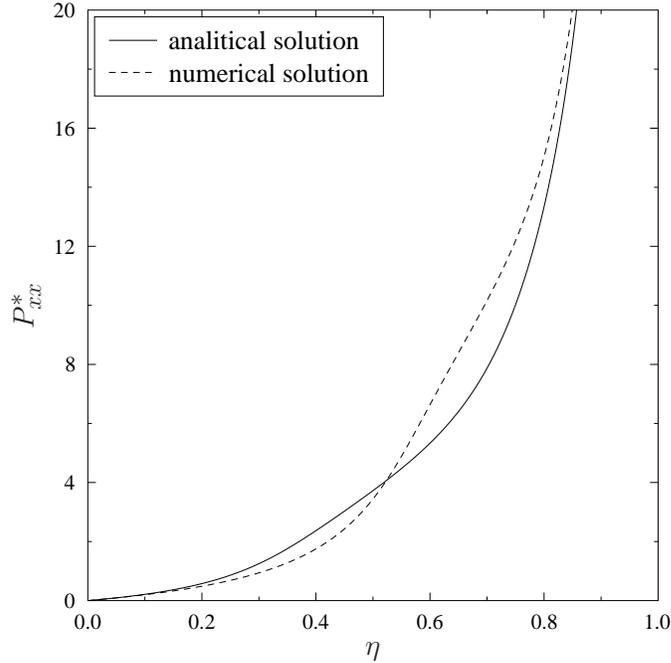}
\end{center}
\caption{Equation of state (longitudinal pressure component vs. packing fraction) of the confined hard disks and that of analytical model where $P^*=\beta P_{xx}WD$ and $\eta=\rho\sqrt{D^2-W^2}$.\label{fig:6}}
\end{figure*}

\begin{figure*}
\begin{center}
\im[width=0.6\textwidth]{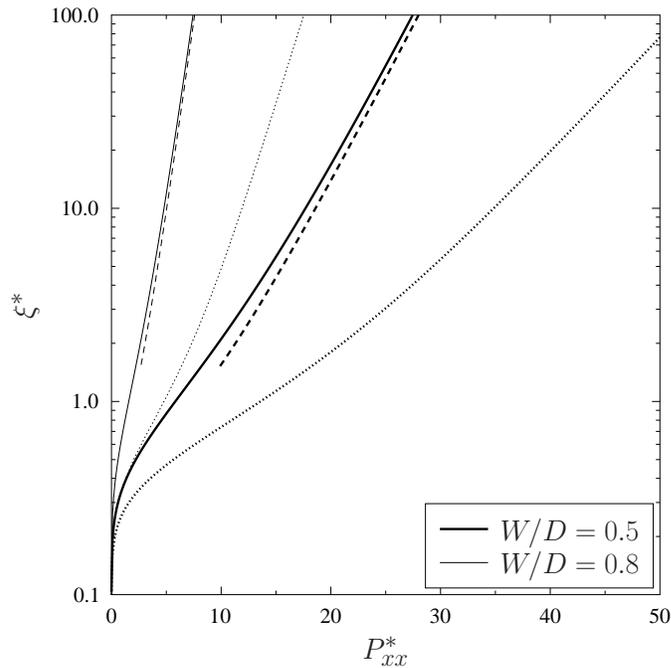}
\end{center}
\caption{Comparison of the correlation lengths of the confined hard disks and that of the analytical model, where $\xi^*=\xi/D$ and $P^*=\beta P_{xx}WD$. Solid curves corresponds to the analytical model, dashed curve to the asymptotic formula for the analytical model and short dashed curve for the hard disks.\label{fig:7}}
\end{figure*}

\begin{figure*}
\begin{center}
\im[width=0.6\textwidth]{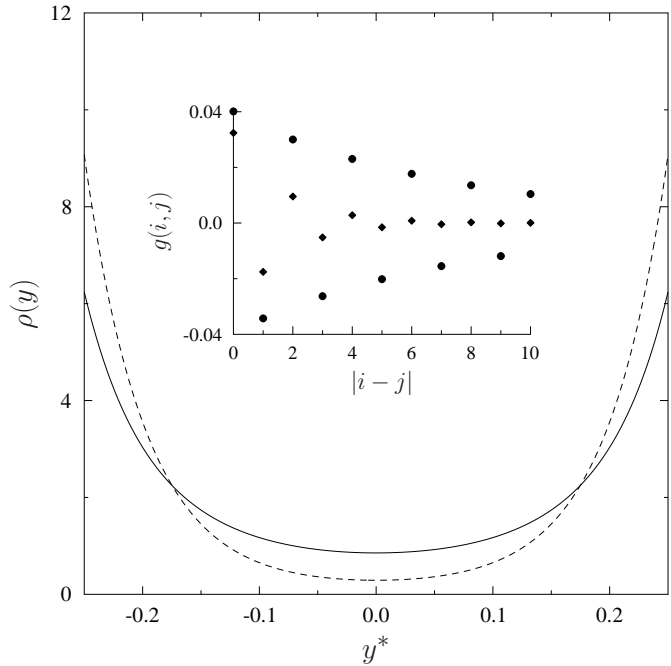}
\end{center}
\caption{Density profile of confined hard disks (solid curve) and that of analytical model (dashed curve) at pore width $W=D/2$ for $\eta=\sqrt{3}/2$. Inset shows the corresponding transverse pair correlation function, where diamonds correspond to the confined hard disks, while the filled circles to the analytical model. \label{fig:8}}
\end{figure*}

In the rest of this section our analytical and the numerical results are compared and we discuss the strength and the weakness of the applied approximation for the contact distance. Figure \ref{fig:6} shows the analytical and the numerical longitudinal pressures as a function of packing fraction for $W=0.8D$. 
We can see that the low and high density behaviours of both solutions agree very well. This is not surprising because both are ideal gas at low density and both are packed in the same alternating (triangular lattice-like) structure at the close packing. However, the two pressure curves deviate substantially at the intermediate pressures and even a crossover takes place between them at $\eta\approx 0.5$.  It is easy to understand that these behaviours stem from the exaggeration of the contact distance for small transverse positional differences. Figure \ref{fig:2} shows that the contact distance is overestimated by about $70\%$ for $|y_i-y_{i+1}|=0$ in the case of $W=0.8D$. Consequently the particles of the analytical model always interact with larger contact distances in all horizontal positions and they occupy larger regions of the space. As a result, the analytical model possess higher pressure in the Tonks gas regime and it undergoes the fluid-solid structural change at lower densities than the system of hard disks. We can also see that the particles are pushed more strongly to the walls in the analytical model than in the numerical one to produce lower pressures at high densities. In the analytical model it is feasible to determine the high density limit of the pressures analytically. We get that $\beta P_{xx}W=2\rho/\Big(1-\rho\sqrt{D^2-W^2}\Big)$ and $P_{yy}/P_{xx}=\frac{1}{2}\Big(1+1/(1-W^2/D^2)\Big)$. The first expression tells us that the structural change gives an extra factor two in the well-known Tonks equations, while the second one shows that the transverse pressure becomes higher than the longitudinal one for all possible widths at the close packing. Note that the above relation between $P_{xx}$ and $P_{yy}$ is probably not true for hard disks and the phase behaviour of the two systems is not the same at high densities. We cannot prove it, because the eigenvalue equation \eref{eq:6} cannot be solved numerically at very high pressures. However the examination of the correlation length reveals the differences residing in the two models (see figure \ref{fig:7}). 
For the analytical model an expression can be derived for very high pressures, which is given by $\xi=-\exp(-c\beta P_{xx}W^2)/(4c\beta P_{xx}W^2)$, while $\xi$ is determined from equations \eref{eq:15} and \eref{eq:16}. Apart from the low-density limit the numerical and the analytical correlation lengths diverges with different slope with increasing pressure. We can see that the analytical model is more correlated than the numerical one at the same pressure. It is also clear that the pressure is sufficiently high for the comparison because the analytical expression for the high pressure limit of the correlation length fits the exact one. The reason for the difference in the high pressure limit of the correlations is that the analytical model can pack more efficiently than the numerical one. The numerical model corresponds to the confined system of hard rhombuses with longer horizontal extension, i.e. it is harder for them to move between the two walls without overlapping with the neighbors. The uncorrelated movement is much easier for the disks, because they are more rounded than the rhombuses. Therefore the first order expansion of the contact distance gives rise to a stronger tendency for positional ordering and the two systems do not behave identically for very high pressures. Finally we compare the density profiles and the correlation functions of the two models in figure \ref{fig:8}. 
It confirms our present results that the analytical model is more ordered than the hard disks at the walls, the correlations are also stronger in the analytical model and the structure of solid-like clusters are triangular lattice-like in both models. 

In summary we can say that apart from some quantitative differences between the two models, it is proved very useful to devise the analytical model because it follows closely the phase behaviour of the confined hard disks. Furthermore analytical equations can be derived for the pressure, density profile and correlation function, which make possible to examine the high pressure limit of the equation of state and the correlation length. 

\section{Conclusions}
\label{sec:concl}
In this paper we have studied the phase behaviour of two-dimensional hard disks confined between two parallel lines by the transfer-matrix method and MC simulations. The width of the tube is restricted to the range of $0<W<D\sqrt{3}/2$ to guarantee the exact statistical mechanical treatment of the system by taking into account the nearest neighbour interactions only. In summary the following important results are obtained:
\begin{enumerate}[(i)]
\item The numerical investigation of the thermodynamic and structural properties has shown that no phase transition occurs in very narrow tubes. However there are some indications of the structural change between a fluid-like and the solid-like structures. These are the followings: 
\begin{enumerate}
\item the longitudinal pressure component may possess two inflection points, 
\item the longitudinal pressure deviates from the ideal gas law and it may exceed the transverse pressure component, 
\item the pair correlation function shows damped oscillation, decays exponentially with the distance and it becomes infinitely long-ranged in the close packing limit,
\item the pair distribution function changes from the typical Tonks gas behaviour to a solid-like one having shorter period.
\end{enumerate}
\item It has been managed to device an exactly solvable system, which is obtained by the first order Taylor expansion of the contact distance between two hard disks. It has been shown that the new system corresponds to the system of parallel rhombuses between parallel walls. The eigenfunctions and eigenvalues and the corresponding pressure components, density distribution and pair correlation function are determined in analytical form. The analytically solvable model reproduces all important properties of the confined hard disks and makes possible to determine the close packing properties of the pressure and the correlation length.
\item The numerical, analytical and simulation studies show together that the disks arrange themselves into an up-down sequence, where the neighboring particles are forced to stay at the opposite walls. The size of the alternating (triangular lattice-like) clusters becomes larger with increasing density and diverges at the close packing density. At such packing where the dominant interactions changes their directions from the horizontal direction to the vertical one, the longitudinal pressure and the pair distribution function shows unexpected behaviours.  
\end{enumerate}
Even though our present study is exact only for very narrow tubes we believe that it is a step forward to understand the crystallization of two-dimensional bulk and confined fluids. It would be especially interesting to go beyond the nearest neighbor restriction and examine the effect of the width on the nature of the structural rearrangement. It may happen that the correlation length diverges at finite density and first order phase transition takes place between a fluid and solid phases. In very narrow pores it may possible to examine the dynamical properties of the hard disk analytically, which may give deeper insight into the triangular lattice solidification of the fluid. One step ahead in this direction is the recent study of Arenzon et. al. \cite{arenzon}, where the dynamical properties of the one-dimensional system of rotating hard rectangles have been examined. In the very recent study of Herrera-Velarde et al. \cite{herrera}, common structural and dynamical behaviors have been found at the transitions from fluid-like to solid-like structures in some repulsive one-dimensional systems.


\end{document}